\documentclass[aps,pre,groupedaddress,twocolumn]{revtex4}

\usepackage{graphicx}
\usepackage{amssymb}
\renewcommand{\vec}[1]{{\mathbf #1}}
\newcommand{\func}[1]{{\mathrm{#1}}}

\begin{document}

\title{Diffuse waves in nonlinear disordered media$^0$}
\author{S.E. Skipetrov}
\email[]{Sergey.Skipetrov@grenoble.cnrs.fr}
\author{R. Maynard}
\affiliation{Laboratoire de Physique et Mod\'elisation des Milieux Condens\'es,
CNRS, 25 Avenue des Martyrs, 38042 Grenoble, France}

\date{\today}
\maketitle

\section{Introduction}
\label{secintro}

The\footnotetext[0]{To appear in ``Wave Scattering in Complex Media: From Theory to Applications'',
edited by B.A. van Tiggelen and S.E. Skipetrov (Kluwer Academic Publishers, 2003).} field of multiple scattering of classical waves
(electromagnetic, acoustic and elastic waves, etc.) in
disordered media has revived in the eighties
\cite{albada85}--\cite{sheng90}, when the
far-reaching analogies between the diffuse transport of waves
and electrons in mesoscopic systems have been realized
(see Ref.\ \cite{tiggelen94} for a comprehensive review of the latter issue).
Using classical waves to study such
phenomena as weak and
strong localization,
mesoscopic correlations and universal conductance
fluctuations
(see Refs.\ \cite{berk94}--\cite{sebbah01} for reviews)
appears to be advantageous in many
aspects: no need for low
temperatures and small samples, better control of the experimental apparatus,
possibility of more sensitive measurements, etc.
In addition, experiments
with classical waves can be readily performed in the {\em linear\/} regime,
excluding interaction between scattered waves and hence simplifying
the interpretation of experimental data, whereas the electron-electron
interaction is always present in the realm of mesoscopic electronics
and cannot simply be `turned off', introducing significant
difficulties in the theoretical model \cite{alt91}.

The `interaction' of classical waves, analogous in some sense to the
electron-electron interaction, can come about if the waves propagate in
a {\em nonlinear\/} medium. Such an interaction is a matter of scientific enquiry
in the fields of nonlinear optics \cite{boyd02}, nonlinear acoustics \cite{hamilton78},
etc. Nowadays, the latter are
rapidly developing scientific disciplines on their own right, having
considerable fundamental importance and numerous practical applications.
Various nonlinear phenomena (self-action: self-phase modulation and self-focusing
of pulses and beams, harmonic generation, shock wave
formation, etc.) occur in {\em homogeneous\/} nonlinear
media depending on the type and strength of the nonlinearity.
The effect of {\em weak\/} disorder
on the propagation of nonlinear waves has also been studied both
theoretically (treating it as a weak perturbation)
and experimentally (e.g., for laser beam propagation through atmospheric
turbulence) \cite{kandidov96}.
Unfortunately, only few experiments \cite{boer93} have been performed up to now
on multiple, {\em diffuse\/} scattering of classical waves in nonlinear media.
Theoretical analysis of the problem is more advanced (see the bibliography of
Ref.\ \cite{skip00}), but is still
insufficient to stimulate further experimental effort.
Meanwhile, diffuse waves seem to be
a good candidate for a detailed study
of the combined effect of disorder and nonlinearity (or interaction) on wave
(or quantum particle) propagation, as
compared to interacting electrons in disordered mesoscopic samples:
the strength and the type of nonlinear wave `interaction' can be readily controlled
and the nonlinear term in the wave equation is often of simple algebraic
form. Despite these important simplifications, diffusion of classical waves
in nonlinear media remains an involved problem and is far from being solved.
The purpose of the present paper is to review our recent
theoretical results \cite{skip00}--\cite{skip02b} which are particularly
susceptible to stimulate the reader's interest, further theoretical analysis,
and, hopefully, new experiments.

For the sake of concreteness,
we restrict our consideration to the problem of self-action of a scalar monochromatic
wave (frequency $\omega$) in a nonlinear disordered medium. This is
described by the following wave equation:
\begin{eqnarray}
&&\left\{ \nabla^2 -
\frac{1}{c^2} \frac{\partial^2}{\partial t^2} \left[
1 + \delta\varepsilon(\vec{r}) +
\Delta\varepsilon_{\mathrm{NL}}(\vec{r}, t) \right]
\right\} E(\vec{r}, t)
\nonumber \\
&&\hspace*{2cm} = J(\vec{r}, t),
\label{weq}
\end{eqnarray}
where $E(\vec{r}, t)$ is the wave amplitude,
$c$ is the speed of the wave in the average medium,
$J(\vec{r}, t) = J_0(\vec{r})$ $\exp(-i \omega t)$
is a monochromatic source term,
$\delta\varepsilon(\vec{r})$
is the fractional fluctuation of the linear dielectric constant,
and the nonlinear part of the dielectric constant
$\Delta\varepsilon_{\mathrm{NL}}(\vec{r}, t)$ depends
on the intensity $I(\vec{r}, t) = \left| E(\vec{r}, t) \right|^2$ of the wave.
Eq.\ (\ref{weq}) describes, e.g., propagation of light in a medium with
intensity-dependent refractive index and (as the reader might already have noted)
we adopt the `optical' terminology from here on. Furthermore, we consider
$E$ in Eq.\ (\ref{weq}) to be a complex quantity, thus neglecting the (possible)
generation of the third harmonics, and restrict ourselves to a weak nonlinearity
(see below for the definition of `weakness').

There are two fundamental questions to consider concerning the propagation
of waves in a disordered
medium with a weak nonlinearity. First, one can ask about the effect of
nonlinearity on the phenomena known for waves
in linear disordered media.
We partially answer this question
in Sec.\ \ref{secangular}, where we summarize the results of calculation
of the angular correlation functions of scattered waves and the analysis of the coherent
backscattering cone in a nonlinear disordered medium.
The second question is more challenging: can the weak nonlinearity give rise to
new physical phenomena which are not present in the linear medium? The answer to the latter question
is `yes' as we demonstrate in Sec.\ \ref{secinst} where the
instability of diffuse, multiple-scattered waves in a disordered medium with
a weak nonlinearity is considered.
In order to facilitate the presentation of our main results, we start by a very brief
review of principal results known for waves in linear disordered 
(Sec.\ \ref{seclinear}) and homogeneous nonlinear (Sec.\ \ref{secnonlinear})
media.

\section{Waves in Linear Disordered Media}
\label{seclinear}

Waves propagating in a {\it linear\/} disordered medium are described by
Eq.\ (\ref{weq}) with $\Delta\varepsilon_{\mathrm{NL}}(\vec{r}, t) = 0$.
Despite its linearity, this problem is a rather involved one and it
attracts a lot of
attention (see, e.g., the articles in the present book and Refs.\ \cite{rossum99}
and \cite{sebbah01}).
In what follows, we are
interested uniquely in the diffusion regime of wave propagation that is
realized for $k_0 \ell \gg 1$, where $k_0 = \omega/c$ and $\ell$ is
the mean free path due to disorder. To simplify still further
the consideration, we assume that
the correlation length of $\delta\varepsilon(\vec{r})$
is much shorter than the
wavelength $\lambda$ and adopt the model of the Gaussian white-noise
disorder:
$\left< \delta\varepsilon(\vec{r}) \delta\varepsilon(\vec{r}_1) \right> =
4 \pi/(k_0^4 \ell) \delta(\vec{r} - \vec{r}_1)$, so that the scattering and
the transport mean free paths coincide.
Besides, we consider a disordered medium without absorption
(i.e. both $\delta\varepsilon$ and $\Delta\varepsilon_{\mathrm{NL}}$ are real), except
if the opposite is specified explicitly.
In the regions of space located at distances greater than $\ell$ from the sources of
waves and the boundaries of the sample,
the average intensity $\left< I(\vec{r}) \right>$ then obeys a diffusion equation
\cite{ishim78}:
\begin{eqnarray}
\frac{\partial}{\partial t} \left< I(\vec{r}, t) \right> =
D \nabla^2 \left< I(\vec{r}, t) \right> + S(\vec{r}, t),
\label{difint}
\end{eqnarray}
where $D = c \ell/3$ is the diffusion constant and $S(\vec{r}, t)$ is the source
term. In the following, we assume a time-independent source term in
Eq.\ (\ref{difint}): $S(\vec{r}, t) = S(\vec{r})$, yielding
$I(\vec{r}, t) = I(\vec{r})$.

The average intensity $\left< I(\vec{r}) \right>$ given by Eq.\ (\ref{difint})
is not sufficient to describe the intensity $I(\vec{r})$ of the wave,
since the latter exhibits large fluctuations:
$\delta I(\vec{r}) = I(\vec{r}) - \left< I(\vec{r}) \right>
\sim \left< I(\vec{r}) \right>$ \cite{ishim78,shapiro86}. The spatial correlation
function of $\delta I(\vec{r}, t)$ contains a short-range
($\Delta r < \ell$) but strong ($\sim 1$) contribution \cite{shapiro86}:
\begin{eqnarray}
\left< \delta I(\vec{r}) \delta I(\vec{r}_1) \right> &=&
\left< I(\vec{r}) \right> \left< I(\vec{r}_1) \right>
\left[ \frac{\sin(k_0 \Delta r)}{k_0 \Delta r} \right]^2
\nonumber \\
&\times& \exp\left( -\frac{\Delta r}{\ell} \right)
\label{short}
\end{eqnarray}
and a long-range ($\Delta r > \ell$) but weak [$\sim (k_0 \ell)^{-2} \ll 1$] contribution \cite{stephen87,zyuzin87}:
\begin{eqnarray}
\left< \delta I(\vec{r}) \delta I(\vec{r}_1) \right> \sim
\frac{c}{k_0^2}
I_0^2 G(\vec{r}, \vec{r}_1) \sim \frac{I_0^2}{k_0^2 \ell \Delta r},
\label{long}
\end{eqnarray}
where $\Delta \vec{r} = \vec{r} - \vec{r}_1$,
$I_0$ is a typical value of $\left< I(\vec{r}) \right>$,
$G(\vec{r}, \vec{r}_1)$ is the Green's function of Eq.\ (\ref{difint}) with
$(\partial/\partial t) = 0$, and the last expression in Eq.\ (\ref{long})
is obtained assuming that
$G(\vec{r}, \vec{r}_1)$ is equal to its value in the infinite medium:
$G(\vec{r}, \vec{r}_1) = G_0(\vec{r}, \vec{r}_1) = (4 \pi D \Delta r)^{-1}$.
The short-range contribution to the intensity correlation function
(\ref{short}) describes the speckle structure of the spatial intensity distribution
in disordered media, while the long-range contribution (\ref{long})
testifies that different speckle spots are correlated (weakly, since
$k_0 \ell \gg 1$).
The correlation functions given by Eqs.\ (\ref{short}) and (\ref{long}) are
commonly referred to as $C^{(1)}$ and $C^{(2)}$, respectively.
Other contributions to the intensity correlation function can exist:
$C^{(3)}$ is a factor $(k_0 \ell)^{-2} \ll 1$ smaller
than $C^{(2)}$ \cite{rossum99} and $C^{(0)} \sim (k_0 \ell)^{-1}$ is non-vanishing only if the
source of waves has the size smaller or of the order of the wavelength
$\lambda$ \cite{shapiro99}. We do not consider the two latter
contributions here.

Short- and long-range correlation functions of the field and intensity fluctuations can be also defined
in the angular domain. For our purposes, it will be sufficient to consider only the short-range correlations
$C^{(1)}$.
If a plane wave is incident on a disordered slab of thickness $L \gg \ell$
(slab surfaces being perpendicular to the $z$-axis)
with a wave vector $\vec{k}_a = \{ \vec{q}_a, k_{az} \}$,
where $\vec{q}_a = \{ k_{ax}, k_{ay} \}$, the field (intensity) of the wave
transmitted through the slab with a wave vector $\vec{k}_b$ is $E(\vec{k}_a, \vec{k}_b)$
[$I(\vec{k}_a, \vec{k}_b) = \left| E(\vec{k}_a, \vec{k}_b) \right|^2$].
If the above experiment is repeated for different incoming and outgoing wave vectors
$\vec{k}_a^{\prime}$ and $\vec{k}_b^{\prime}$, respectively, the normalized correlation function
of scattered fields
\begin{eqnarray}
C(\vec{k}_a, \vec{k}_b; \vec{k}_a^{\prime}, \vec{k}_b^{\prime}) &\equiv&
C_{a b a^{\prime} b^{\prime}}
\nonumber \\
&=& \frac{\left< E(\vec{k}_a, \vec{k}_b) E^*(\vec{k}_a^{\prime}, \vec{k}_b^{\prime} \right>)}{
\left< I(\vec{k}_a, \vec{k}_b) \right>^{1/2} \left<I(\vec{k}_a^{\prime}, \vec{k}_b^{\prime}) \right>^{1/2}}
\label{ca}
\end{eqnarray}
appears to have a rather simple form \cite{berk94}:
\begin{eqnarray}
C_{a b a^{\prime} b^{\prime}} =
\delta_{\Delta \vec{q}_a, \Delta \vec{q}_b} F_{\mathrm{T}}(\Delta q_a L),
\label{ca2}
\end{eqnarray}
where $F_{\mathrm{T}}(x) = x/\sinh x$, $\Delta \vec{q}_a = \vec{q}_a - \vec{q}_a^{\prime}$ (and similarly for $\Delta \vec{q}_b$),
and $\Delta q_a \ell \ll 1$ is assumed. 

The field correlation in reflection, when both incident and scattered waves are
on the same side of the slab, is \cite{berk94,bressoux00}
\begin{eqnarray}
C_{a b a^{\prime} b^{\prime}} \simeq
\delta_{\Delta \vec{q}_a, \Delta \vec{q}_b}
\left[ F_{\mathrm{R}}(\Delta \vec{q}_a) + F_{\mathrm{R}}(\vec{q}_a + \vec{q}_b + \Delta \vec{q}_a) \right],
\label{ca3}
\end{eqnarray}
where
\begin{eqnarray}
F_{\mathrm{R}}(\vec{q}) \simeq (1 + 2 \zeta) - 2(1+\zeta)^3 q \ell + \ldots,
\label{fr}
\end{eqnarray}
$\zeta \sim 1$ is the extrapolation factor ($\zeta=2/3$ in the diffusion approximation),
$q \ell \ll 1$, and the semi-infinite disordered medium is considered.
The short-range correlation function of intensity fluctuations, $C^{(1)}$, in transmission (reflection) geometry
is obtained by taking the square of the absolute value of Eq.\ (\ref{ca2}) [Eq.\ (\ref{ca3})].

Finally, a well-studied phenomenon in the linear disordered medium is the coherent backscattering \cite{albada85},
consisting in a two-fold  (with respect to the incoherent, diffuse background)
enhancement of the average scattered intensity in the direction of
exact backscattering ($\vec{k}_b = -\vec{k}_a$). For normal
incidence ($\vec{q}_a = 0$), the angular line shape of the
coherent backscattering cone can be approximately expressed using the function $F_{\mathrm{R}}$ defined in Eq.\ (\ref{fr}):
\begin{eqnarray}
\frac{\left< I(\vec{q}) \right>}{\left< I_{\mathrm{inc}} \right>}
\simeq
\frac{F_{\mathrm{R}}(1/L_{\mathrm{a}}) + F_{\mathrm{R}}[ (q^2 + 1/L_{\mathrm{a}}^2)^{1/2} ]}{
F_{\mathrm{R}}(1/L_{\mathrm{a}})},
\label{cbs}
\end{eqnarray}
where $\vec{q} \equiv \vec{q}_b$ and we have taken into account
the absorption of light in the medium by introducing the macroscopic
absorption length $L_{\mathrm{a}} \gg \ell$.

\section{Waves in Homogeneous Nonlinear Media}
\label{secnonlinear}

If $\delta\varepsilon(\vec{r}, t) = 0$, Eq.\ (\ref{weq}) describes the
propagation of waves in a {\it homogeneous\/} nonlinear medium (no scattering) with
an intensity-dependent dielectric constant \cite{boyd02}.
Below we consider the case of Kerr nonlinearity.
If the nonlinear response of the medium is instantaneous and
local, the dependence of $\Delta\varepsilon_{\mathrm{NL}}(\vec{r}, t)$
on $I(\vec{r}, t)$ is rather simple:
$\Delta\varepsilon_{\mathrm{NL}}(\vec{r}, t) = \varepsilon_2 I(\vec{r}, t)$, where
$\varepsilon_2$ is a nonlinear coefficient. In general, however,
the nonlinear response is not instantaneous and
can be modeled by a phenomenological equation of Debye type for
$\Delta\varepsilon_{\mathrm{NL}}$:
\begin{eqnarray}
\tau_{\mathrm{NL}} \frac{\partial}{\partial t} \Delta\varepsilon_{\mathrm{NL}}(\vec{r}, t) =
-\Delta\varepsilon_{\mathrm{NL}}(\vec{r}, t) + \varepsilon_2 I(\vec{r}, t),
\label{debye}
\end{eqnarray}
where $\tau_{\mathrm{NL}}$ is the response time of the nonlinearity.
Moreover, the non-locality of the nonlinear response of the medium can be taken into
account by considering $\Delta\varepsilon_{\mathrm{NL}}(\vec{r}, t)$ obeying an
equation of diffusion type:
\begin{eqnarray}
\tau_{\mathrm{NL}} \frac{\partial}{\partial t} \Delta\varepsilon_{\mathrm{NL}}(\vec{r}, t) &=&
(a_{\mathrm{NL}})^2 \nabla^2 \left[ \Delta\varepsilon_{\mathrm{NL}}(\vec{r}, t) \right]
\nonumber \\
&-&\Delta\varepsilon_{\mathrm{NL}}(\vec{r}, t) + \varepsilon_2 I(\vec{r}, t),
\label{dif}
\end{eqnarray}
where $a_{\mathrm{NL}}$ is some characteristic length, describing the
degree of non-locality of the nonlinear response.

The physics of nonlinear optical phenomena in Kerr media
is rather rich \cite{boyd02} and it is not our
purpose to review it here.
In the framework of our study, we would like, however, to
call the attention of the reader to the following
two simple but illustrative examples.
First, if a nonlinear Kerr medium, illuminated by a monochromatic plane wave of
intensity $I_0$,
is put in an optical
resonator, the feedback provided by the resonator leads to the instability of
the steady-state solution $E(\vec{r}, t) = E_0(\vec{r}) \exp(-i \omega t)$ and
the intensity $I = \left| E \right|^2$ of the wave develops a time dependence
\cite{ikeda80}.
This phenomenon occurs only at $I_0$ exceeding some threshold
and the dynamics of $I(\vec{r}, t)$ becomes progressively more complex as
$I_0$ increases.
The second example concerns two counter-propagating light
waves in a Kerr medium. Again, the steady-state solution develops an instability
for the intensities of the waves exceeding a threshold \cite{silber82}.
The above simple examples suggest that the unstable regimes are fundamental
in nonlinear optics and this appears to be indeed true:
instabilities, self-oscillations, pattern formation, and spatio-temporal
chaos are observable in many nonlinear optical systems (see, e.g.,
Refs.\ \cite{gibbs85}--\cite{voron99} for reviews).
Even incoherent light beams have been recently shown to
exhibit pattern formation and `optical turbulence'
due to the so-called modulation instability \cite{soljacic00}, indicating that unstable behavior does
not require the perfect coherence of the underlying wave field.

\section{Angular Correlations and Coherent Backscattering for Waves in Nonlinear Disordered Media}
\label{secangular}

After having briefly reviewed the key features of wave propagation in linear disordered and
homogeneous nonlinear media, we are now in a position to attack
the central problem of the present paper: diffuse wave propagation in nonlinear disordered media.
A particularly pictorial description of the latter can be developed
if the nonlinearity is assumed to be weak enough to have no significant effect
on the general diffuse character of wave propagation in the medium and
the values of the diffusion constant $D$ and the mean free path $\ell$.
This requires the `nonlinear scattering length' $\ell_{\mathrm{NL}}$ due
to the nonlinear term in Eq.\ (\ref{weq}) to be much larger than
$\ell$. $\ell_{\mathrm{NL}}$ can be estimated
by assuming the short-range intensity correlation function in a nonlinear
disordered medium to be approximately the same is in the
linear one [see Eq.\ (\ref{short})].
Calculating the total scattering crossection of statistically isotropic fluctuations of the
nonlinear dielectric constant $\varepsilon_2 \delta I(\vec{r})$ in the Born approximation,
we obtain
\begin{eqnarray}
\ell_{\mathrm{NL}}^{-1} \propto k_0^2 \int_0^{2 k_0} \Phi(\kappa) \kappa d \kappa,
\label{ellnl}
\end{eqnarray}
where
\begin{eqnarray}
\Phi(\vec{K}) \propto \varepsilon_2^2 \int \left< \delta I(\vec{r}) \delta I(\vec{r}+
\Delta \vec{r}) \right> \exp(-i \vec{K} \Delta \vec{r}) d^3 \Delta \vec{r},
\label{phiellnl}
\end{eqnarray}
and hence $\ell_{\mathrm{NL}} \sim (\Delta n^2 k_0)^{-1}$, where
$\Delta n = n_2 I_0$ is the typical value of the nonlinear
correction to the refractive index, $n_2 = \varepsilon_2/2$, and
$I_0$ is the typical value of $\left< I(\vec{r}, t) \right>$.
The condition of weak nonlinearity, $\ell \ll \ell_{\mathrm{NL}}$, then becomes
\begin{eqnarray}
\Delta n^2 k_0 \ell \ll 1,
\label{weak}
\end{eqnarray}
where, we recall, $k_0  \ell \gg 1$.
We adopt this condition throughout the rest of the paper.

We start the analysis of the effect of nonlinearity on multiple scattering of waves in disordered media
by considering the angular correlation of scattered wave fields,
$C_{a b a^{\prime} b^{\prime}}$, as defined in Eq.\ (\ref{ca}) (Sec.\ \ref{seclinear}),
assuming the two incident waves to have different amplitudes ($A$ for the wave with the wave vector
$\vec{k}_a$ and $A^{\prime}$ for the wave with the wave vector
$\vec{k}_a^{\prime}$). Due to the nonlinearity of the medium,
$C_{a b a^{\prime} b^{\prime}}$ will now depend not only on
$\vec{k}_a$, $\vec{k}_a^{\prime}$, $\vec{k}_b$, and $\vec{k}_b^{\prime}$, but also on $A$ and $A^{\prime}$.
For simplicity, we put 
$A^{\prime} \rightarrow 0$ (so that the effect of the nonlinearity on the propagation of the
second wave is negligible) and limit ourselves
to the case of instantaneous ($\tau_{\mathrm{NL}} = 0$) local ($a_{\mathrm{NL}} = 0$) nonlinearity.

The physical origin of the nonlinearity-induced modification of $C_{a b a^{\prime} b^{\prime}}$ can be
easily understood in the framework of the path-integral picture of wave propagation.
The wave of amplitude $A$, propagating along some wave path of length $s$ through a nonlinear disordered 
medium, acquires an additional, `nonlinear' phase shift $\Delta \phi_{\mathrm{NL}}$
with respect to the wave of amplitude
$A^{\prime} \rightarrow 0$, propagating in an essentially linear medium:
\begin{eqnarray}
\Delta \phi_{\mathrm{NL}} = k_0 n_2 \int_0^{s} ds_1
I(\vec{r}_1),
\label{phasenl1}
\end{eqnarray}
the average value of the latter being
\begin{eqnarray}
\left< \Delta \phi_{\mathrm{NL}} \right> \approx k_0 n_2 A^2 s,
\label{phasenl2}
\end{eqnarray}
where we assume $\left< I(\vec{r}_1) \right> \approx A^2$. 
In a linear medium,
a similar phase difference arises between two waves at different frequencies $\omega \ne \omega^{\prime}$:
\begin{eqnarray}
\Delta \phi_{\Delta \omega} = (\Delta \omega/c) s,
\label{cw}
\end{eqnarray}
where $\Delta \omega = \omega - \omega^{\prime} \ll \omega$.
It is well-known that the latter dephasing modifies the angular correlation function of scattered fields,
which becomes (in transmission) \cite{berk94}
\begin{eqnarray}
C_{a b a^{\prime} b^{\prime}} =
\delta_{\Delta \vec{q}_a, \Delta \vec{q}_b} 
F_{\mathrm{T}} \left[ (\Delta q_a^2 - 2 i \gamma^2)^{1/2} L \right],
\label{ca4}
\end{eqnarray}
where $\gamma^2 = 3 \Delta \omega/(2 \ell c)$.

\begin{figure}
\begin{center}
{\begin{flushleft}(a)
\end{flushleft}}
\includegraphics[width=6.0cm]{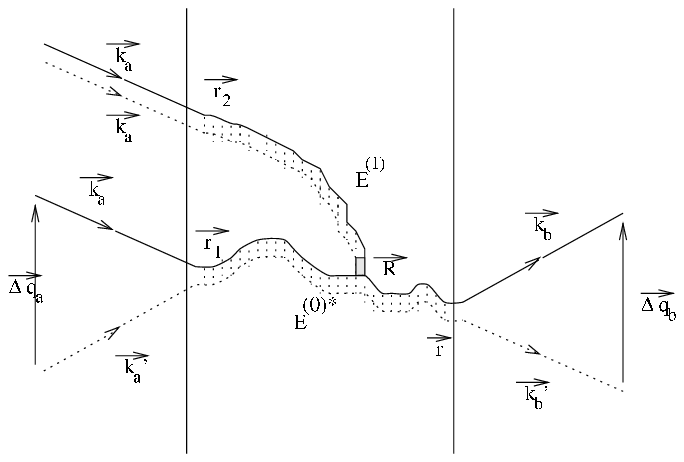}
{\begin{flushleft}(b)
\end{flushleft}}
\includegraphics[width=6.0cm]{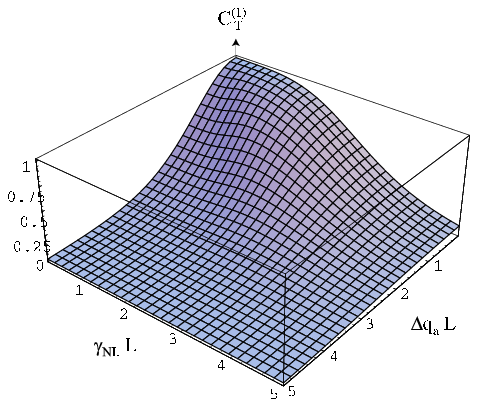}
\caption{\label{fig1}
(a) The diagram contributing to the angular correlation function of scattered wave fields in a
nonlinear disordered medium. (b) Short-range intensity correlation function for a diffuse wave transmitted through
a slab.
The figures are from Ref.\ [17].}
\end{center}
\end{figure}

The formal analogy between Eqs.\ (\ref{phasenl2}) and (\ref{cw}) suggests that as long as the nonlinearity
affects only the phase of the wave and not its amplitude (i.e. for a weak nonlinearity), the correlation function
$C_{a b a^{\prime} b^{\prime}}$ in a nonlinear disordered medium should be approximately given by Eq.\ (\ref{ca4}), where
$\Delta \omega$ is replaced by $\omega n_2 A^2$.
This is indeed confirmed by a perturbative diagrammatic calculation \cite{bressoux00} [see the diagram of Fig.\
\ref{fig1}(a)], that allows one to obtain the lowest term in the series expansion of
$C_{a b a^{\prime} b^{\prime}}$ in $n_2 A^2 \ll 1$.
Applying then a `natural'
ansatz $f(x) - [i \epsilon^2/(2x)] f^{\prime}(x) \simeq f[ (x^2 - i \epsilon^2)^{1/2} ]$
yields \cite{bressoux00}:
\begin{eqnarray}
C_{a b a^{\prime} b^{\prime}} =
\delta_{\Delta \vec{q}_a, \Delta \vec{q}_b} 
F_{\mathrm{T}} \left\{ \left[ \Delta q_a^2 L^2 - 2 i (L/\xi)^2 \right]^{1/2} \right\},
\label{ca5}
\end{eqnarray}
where $\xi$ is a new characteristic length, characterizing the damping of correlation
due to the nonlinear effects:
\begin{eqnarray}
\xi \approx \sqrt{\frac{\ell}{k_0 \Delta n}}.
\label{xi}
\end{eqnarray}
Here $\Delta n \approx n_2 A^2$ is the typical value of the nonlinear part of the
refractive index.
The physical meaning of $\xi $ is clear:
it is the characteristic length for the loss of phase coherence between two waves following
the same diffusion path, one with a vanishing amplitude $A^{\prime} \rightarrow 0$ and the other one with
a finite amplitude $A$.
By analogy with the case of frequency correlation [Eq.\ (\ref{ca4})],
$\gamma_{\mathrm{NL}}$ can be defined as $\gamma_{\mathrm{NL}} = 1/\xi$.
The short-range correlation function of intensity fluctuations, $C_{\mathrm{T}}^{(1)}$, can be obtained by
taking a square of the absolute value of Eq.\ (\ref{ca5}). We show this correlation function in Fig.\ \ref{fig1}(b).

\begin{figure}
\begin{center}
{\begin{flushleft} (a)
\end{flushleft}}
\includegraphics[width=6.0cm]{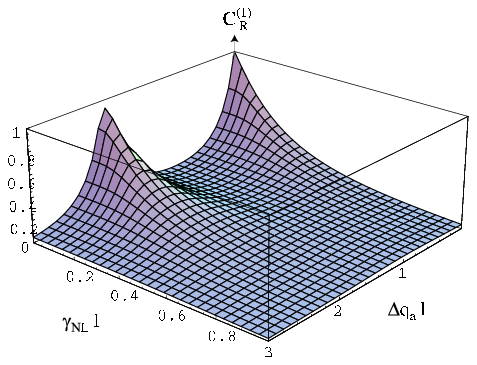}
{\begin{flushleft} (b)
\end{flushleft}}
\includegraphics[width=6.0cm]{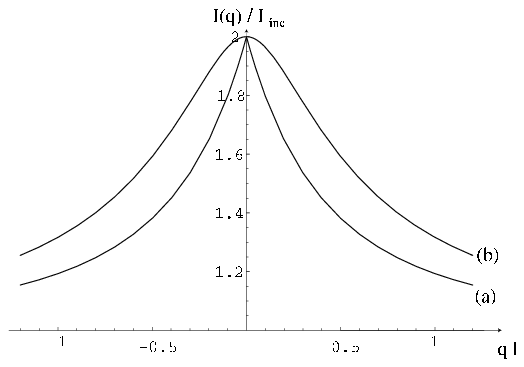}
\caption{\label{fig2}
(a) Short-range angular correlation function of the fluctuations of diffusely reflected intensity in a semi-infinite nonlinear
disordered medium. (b) Line shape of the coherent backscattering cone in a semi-infinite nonlinear disordered medium
for $L_{\mathrm{a}}^{\mathrm{NL}} \rightarrow \infty$ [vanishing linear and nonlinear absorption, curve (a)] and
$L_{\mathrm{a}}^{\mathrm{NL}} = 10 \ell$ [curve (b)]. The figures are from Ref.\ [17].}
\end{center}
\end{figure}

Similarly to the case of transmission geometry, correlation between diffusely reflected waves can also be
calculated taking into account the additional dephasing due to the nonlinearity of the medium.
Again, assuming that the two incident waves have different amplitudes
$A$ and $A^{\prime} \rightarrow 0$, one finds for a semi-infinite disordered medium \cite{bressoux00}
\begin{eqnarray}
&&C_{a b a^{\prime} b^{\prime}} \simeq
\delta_{\Delta \vec{q}_a, \Delta \vec{q}_b}
\left\{ F_{\mathrm{R}} \left[ \left( \Delta \vec{q}_a^2 - 2 i (1+\zeta)/\xi^2 \right)^{1/2} \right] \right.
\nonumber \\
&&+ \left.
F_{\mathrm{R}} \left[ \left( \left| \vec{q}_a + \vec{q}_b + \Delta \vec{q}_a \right|^2 -
2 i (1+\zeta)/\xi^2 \right)^{1/2} \right]
\right\}.\hspace*{6mm}
\label{ca6}
\end{eqnarray}
This result is to be compared to the `linear' result (\ref{ca3}).
In Fig.\ \ref{fig2}(a) we show the experimentally accessible short-range
correlation function of intensity fluctuations, $C_{\mathrm{R}}^{(1)}$,
obtained by taking a square of the absolute value of Eq.\ (\ref{ca6}).
Just as in a linear medium, two peaks occur at $\Delta \vec{q}_a = 0$ and $\vec{q}_a + \vec{q}_b + \Delta \vec{q}_a = 0$,
respectively.

Finally, the coherent backscattering cone can be calculated at the same level of approximation
\cite{bressoux00}.
If the nonlinear coefficient $\varepsilon_2$ in Eq.\ (\ref{weq}) is purely real, no deviation from the linear
result (\ref{cbs}) is found within the present theoretical framework.
The contribution of the crossed diagrams to the intensity of backscattered wave is, however, modified
if $\varepsilon_2$ has a small imaginary part (i.e. in the presence of nonlinear absorption).
This phenomenon can be described by introducing a
generalized absorption length $L_{\mathrm{a}}^{\mathrm{NL}}$, accounting for both linear and nonlinear absorption:
\begin{eqnarray}
\frac{1}{(L_{\mathrm{a}}^{\mathrm{NL}})^2} \approx  \frac{1}{L_{\mathrm{a}}^2} + 
\frac{k_0}{\ell} (1 + \zeta) A^2 \func{Im} \varepsilon_2.
\label{lastar}
\end{eqnarray}
Additional absorption, introduced by the nonlinear term in Eq.\ (\ref{weq}), will lead
to the rounding of the triangular peak shape of the coherent backscattering cone, as illustrated in
Fig.\ \ref{fig2}(b).
Mathematically, this phenomenon is described by Eq.\ (\ref{cbs}) with $L_{\mathrm{a}}$ replaced
by $L_{\mathrm{a}}^{\mathrm{NL}}$.

To conclude this section, we would like to emphasize that in calculating the angular correlation
functions [Eqs.\ (\ref{ca5}), (\ref{ca6}) and Figs.\ \ref{fig1}(b), \ref{fig2}(a)]
and the coherent backscattering cone [Fig.\ \ref{fig2}(b)], we have completely neglected
the \textit{fluctuations} of the wave intensity inside the disordered medium, replacing
$I(\vec{r})$ by $\left< I(\vec{r}) \right> \approx A^2 = \mathrm{const}$ in the very beginning
[Eq.\ (\ref{phasenl2})].
The intensity fluctuations in disordered media are, however, known to be important and exhibit some
nontrivial features [such as, e.g., long-range correlations, see Eq.\ (\ref{long})]. It is therefore important to
analyze the role of these fluctuations in the presence of nonlinearity, when the highly irregular
spatial intensity  distribution (speckle pattern) can give rise to a random modification
of the dielectric constant (and, consequently, refractive index).
Such an analysis is presented in the following section.

\section{Instability of Diffuse Waves in Nonlinear Disordered Media}
\label{secinst}

As we already mentioned in Sec.\ \ref{secnonlinear}, unstable regimes are encountered
in many nonlinear optical systems. It is therefore natural to expect
that they may appear
in a disordered nonlinear medium as well. The origin of the
instability in this case is relatively easy to understand in the framework
of the path-integral picture of wave propagation \cite{skip00}.

\subsection{Heuristic description}
\label{ssecheur}

In the path-integral picture, the diffusely scattered wave field at some position
$\vec{r}$ inside the disordered medium is represented
as a sum of partial waves traveling along various diffuse paths.
Due to the nonlinearity of the medium, the phase $\phi(\vec{r}, t)$
of a given partial wave arriving at $\vec{r}$ at time $t$ depends
on the spatio-temporal
distribution of intensity $I(\vec{r}, t)$ (speckle pattern) inside the
medium. At the same time, $I(\vec{r}, t)$ is a result of interference of many
partial waves, and, therefore, $I(\vec{r}, t)$ is sensitive to their phases $\phi(\vec{r}, t)$.
This leads to a sort of feedback mechanism: The phases of partial waves depend
on the speckle pattern $I(\vec{r}, t)$, while the latter depends on the
phases. It appears that such a feedback can destabilize the time-independent speckle pattern $I(\vec{r})$,
leading to $I(\vec{r}, t)$ that fluctuates spontaneously with time, if the nonlinearity is strong enough \cite{skip00}.
As we show below, this phenomenon can be understood in reasonably simple terms
in the case of local ($a_{\mathrm{NL}} = 0$)
and instantaneous ($\tau_{\mathrm{NL}} = 0$) Kerr nonlinearity.

Consider a partial wave traveling along a typical
diffuse path of length $s_0(t) \sim L^2/\ell$ from
the source of waves to some point $\vec{r}$ located at a distance $L$ from
the source. We assume that the path length $s_0(t)$ can vary slowly with time due to, e.g., the motion
of scattering centers [corresponding to a time-dependent disorder $\delta \varepsilon(\vec{r}, t)$ in
Eq.\ (\ref{weq})], while the initial and final points of the path are fixed.
The variations of $s_0$ with time are assumed to be slow (i.e.
$s_0$ does not change significantly during the time $T_{\mathrm{D}} = L^2/D$
required for the wave to cover the distance of order $s_0$).
The phase $\phi(\vec{r}, t)$ of the wave
contains a `linear' contribution
$\phi_{\mathrm{L}}(\vec{r}, t) = k_0 s_0(t)$ and the `nonlinear' one
\begin{eqnarray}
\phi_{\mathrm{NL}}(\vec{r}, t) &=& k_0 \int_0^{s_0(t)} ds_1
[\Delta \varepsilon_{\mathrm{NL}}(\vec{r}_1, t_1)/2]
\nonumber \\
&=& k_0 n_2 \int_0^{s_0(t)} ds_1 I(\vec{r}_1, t_1),
\label{phasenl}
\end{eqnarray}
where $\Delta \varepsilon_{\mathrm{NL}}(\vec{r}_1, t_1)/2
= n_2 I(\vec{r}_1, t_1)$ is the nonlinear part of the
refractive index, $s_1$ is a curvilinear coordinate of the point $\vec{r}_1$
along the path, $t_1 = t - [s_0(t)-s_1]/c$ is the time of wave passage through
$\vec{r}_1$, and the integrals are along the path.

Let the fluctuating part of the linear
dielectric constant $\delta \varepsilon(\vec{r}, t)$ to change
slightly during some time interval $\Delta t$:
$\delta \varepsilon(\vec{r}, t) \rightarrow
\delta \varepsilon(\vec{r}, t + \Delta t) = \delta \varepsilon(\vec{r}, t) +
\Delta [\delta \varepsilon(\vec{r}, t)]$, leading to corresponding changes
$\Delta I(\vec{r}, t, \Delta t)$ and $\Delta \phi(\vec{r}, t, \Delta t)$ of
intensity and phase, respectively.
The latter can be, in principle, found from Eq.\ (\ref{weq}).
$\Delta \phi$ contains
the linear and nonlinear parts,
$\Delta \phi_{\mathrm{L}}$ and $\Delta \phi_{\mathrm{NL}}$, respectively.
Under very general assumptions, one can show that
$\left< \Delta I \right> = 0$ and $\left< \Delta \phi \right> = 0$,
where the averaging is over  $\delta \varepsilon(\vec{r}, t)$,
$\Delta [\delta \varepsilon(\vec{r}, t)]$, and over all paths of the same length $s_0$.
The second moment
$\left< \Delta \phi^2 \right> = \left< \Delta \phi_{\mathrm{L}}^2 \right> +
\left< \Delta \phi_{\mathrm{NL}}^2 \right> > 0$, where
$\left< \Delta \phi_{\mathrm{L}}^2 \right>$ is calculated
in the framework of the so-called diffusing-wave spectroscopy (DWS) \cite{maret87} and
depends on the way in which $\delta \varepsilon(\vec{r}, t)$ is modified. If, for
example, the disordered medium is a suspension of Brownian particles
(particle diffusion coefficient $D_{\mathrm{B}}$),
$\left< \Delta \phi_{\mathrm{L}}^2 \right> = (\Delta t/\tau_0) (s_0/\ell)$,
where
$\tau_0 = (4 k_0^2 D_{\mathrm{B}})^{-1}$ \cite{maret87}.
For the variance of the nonlinear phase difference,
Eq.\ (\ref{phasenl}) yields
\begin{eqnarray}
&&\left< \Delta \phi_{\mathrm{NL}}^2 \right> =
k_0^2 n_2^2 \int_0^{s_0} ds_1 \int_0^{s_0} ds_2
\nonumber \\
&&\hspace*{1cm} \left< \Delta I(\vec{r}_1, t_1) \Delta I(\vec{r}_2, t_2) \right>,
\label{varnl}
\end{eqnarray}
where $t_2 = t_1 + \Delta t$ and both integrals are along the same diffusion path.
Obviously,
\begin{eqnarray}
&&\left< \Delta I(\vec{r}_1, t_1) \Delta I(\vec{r}_2, t_2) \right> =
2 \left[ \left< \delta I(\vec{r}_1, t) \delta I(\vec{r}_2, t) \right> \right.
\nonumber \\
&&\hspace*{1cm}- \left.\left< \delta I(\vec{r}_1, t) \delta I(\vec{r}_2, t + \Delta t) \right> \right]
\nonumber \\
&& \hspace*{1cm} \simeq \left< \Delta \phi_{\mathrm{L}}^2 \right>
\left< \delta I(\vec{r}_1, t) \delta I(\vec{r}_2, t) \right>,
\label{vardi}
\end{eqnarray}
where the second line is obtained by assuming that
$\left< \delta I(\vec{r}_1, t) \delta I(\vec{r}_2, t + \Delta t) \right>$
in the nonlinear
medium is close to its value in the linear one (a sort of perturbation theory),
replacing $\left< \delta I(\vec{r}_1, t) \delta I(\vec{r}_2, t + \Delta t) \right>$
by the `linear' result:
\begin{eqnarray}
&&\left< \delta I(\vec{r}_1, t) \delta I(\vec{r}_2, t + \Delta t) \right>
\simeq \left< \delta I(\vec{r}_1, t) \delta I(\vec{r}_2, t) \right>
\nonumber \\
&& \hspace*{1cm}= \exp\left[-(1/2) \left< \Delta \phi_{\mathrm{L}}^2 \right> \right],
\label{vardi2}
\end{eqnarray}
and assuming $\left< \Delta \phi_{\mathrm{L}}^2 \right> \ll 1$.

Substituting Eq.\ (\ref{vardi}) into Eq.\ (\ref{varnl}) and changing the variables
of integration to $s = (s_1 + s_2)/2$ and $\Delta s = s_1 - s_2$, we obtain
\begin{eqnarray}
\left< \Delta \phi_{\mathrm{NL}}^2 \right> =
\left< \Delta \phi_{\mathrm{L}}^2 \right>
k_0^2 n_2^2 \int_0^{s_0} ds \int_{-s}^{s} d \Delta s
\left< \delta I(\vec{r}_1) \delta I(\vec{r}_2) \right>.
\nonumber \\
\label{varnl2}
\end{eqnarray}
As we discussed in Sec.\ \ref{seclinear},
the correlation function of intensity fluctuations
$\left< \delta I(\vec{r}_1) \delta I(\vec{r}_2) \right>$ in Eq.\ (\ref{varnl2})
contains two contributions: a short-range one [Eq.\ (\ref{short})]
and a long-range one [Eq.\ (\ref{long})]. 
To perform the integration in Eq.\ (\ref{varnl2}), we replace $\Delta r = \left| \vec{r}_1 - \vec{r}_2 \right|$
in the expression (\ref{short}) for the short-range correlation function by $\Delta s$, assuming the
wave path to be ballistic at distances shorter than $\ell$.
In contrast, for $\Delta r > \ell$ the wave path is diffusive, and hence $\Delta r$ in the expression
(\ref{long}) for the long-range correlation function can be substituted by 
$(\Delta s \ell)^{1/2}$.  
Performing then integrations in Eq.\ (\ref{varnl2}) yields
\begin{eqnarray}
\left< \Delta \phi_{\mathrm{NL}}^2 \right> \simeq
p \left< \Delta \phi_{\mathrm{L}}^2 \right>,
\label{varnl3}
\end{eqnarray}
where we introduce the {\em bifurcation parameter} \cite{skip00}
\begin{eqnarray}
p = \Delta n^2 \left( \frac{L}{\ell} \right)^2
\left( k_0 \ell + \frac{L}{\ell} \right),
\label{p}
\end{eqnarray}
and the numerical factors of order unity are omitted.

Note that Eq.\ (\ref{p}) is a sum of two contributions. The first one,
$\Delta n^2$ $( L/\ell )^2 k_0 \ell$, originates from the short-range correlation
of intensity fluctuations [Eq.\ (\ref{short})], while the second contribution,
$\Delta n^2 ( L/\ell)^3$, is due to the long-range correlation [Eq.\ (\ref{long})].
If $p \ll 1$, the nonlinear term
$\left< \Delta \phi_{\mathrm{NL}}^2 \right>$ represents just a small correction
to the linear one $\left< \Delta \phi_{\mathrm{L}}^2 \right>$ and our
perturbation approach is likely to be valid.
In contrast, the above perturbation theory diverges for $p > 1$, since
$\left< \Delta \phi_{\mathrm{NL}}^2 \right>$ becomes larger than
$\left< \Delta \phi_{\mathrm{L}}^2 \right>$. This suggests that
$p \simeq 1$ is a critical point beyond which (i.e. for $p > 1$)
the physics of the nonlinear problem is no longer similar to the physics of the
linear one and hence the perturbation theory cannot be used.
Although the mere breakdown of the perturbation theory does not permit to
draw any far-reaching conclusions
about the speckle pattern beyond the critical point $p \simeq 1$,
a more rigorous analysis (see Secs.\ \ref{ssecpath} and \ref{sseclang})
shows that $p \simeq 1$ defines the instability threshold of the
multiple-scattering speckle pattern, and that at $p > 1$
the latter should exhibit spontaneous
temporal fluctuations even for a time-independent disorder
$\delta \varepsilon(\vec{r})$ (i.e. for vanishing mobility of scattering centers:
$D_{\mathrm{B}} \rightarrow 0$ and $\tau_0 \rightarrow \infty$ in the case of Brownian motion).

It is pertinent to note the extensive nature of the bifurcation
parameter $p$. According to Eq.\ (\ref{p}),
weak nonlinearity can be efficiently compensated by a sufficiently
large sample size $L$ and $p \sim 1$ can be reached even at vanishing
$\Delta n$, provided that $L$ is large enough. In the limit of
large sample size $L \gg k_0 \ell^2$, one gets $p \simeq \Delta n^2 ( L/\ell)^3$
(as also found in Ref.\ \cite{spivak00}). This result is due to the long-range
correlation (\ref{long}) of intensity fluctuations which dominates
in the integral of Eq.\ (\ref{varnl2}) because of its slow decrease
($\propto \left| \vec{r}_1 - \vec{r}_2 \right|^{-1}$)
and consequent extension over the whole disordered sample.
Another remarkable feature of Eq.\ (\ref{p}) is the independence of $p$ of
the sign of $\Delta n$. The phenomenon of the speckle pattern instability is
therefore expected to develop in a similar way for both positive and
negative nonlinear coefficients $n_2$. This is not common for the instabilities
in nonlinear systems without disorder \cite{ikeda80}--\cite{soljacic00} since the latter
are often related to the self-focusing effect, arising at $n_2 > 0$ only. The instability
discussed in the present paper is of different type and has nothing to do with
the self-focusing [which is negligible if the condition (\ref{weak}) is satisfied].
This does not mean, however, that the phenomena similar to those discussed
in the present paper
do not exist in homogeneous nonlinear media. In fact, it is easy to see that
Eq.\ (\ref{phasenl}) describes nothing but
the {\em self-phase modulation\/} \cite{boyd02}
of the multiple-scattering speckle pattern.
It is worth noting that the development of the self-phase
modulation in a disordered medium appears to be rather different from that in the
homogeneous case. Indeed, in a homogeneous medium of size $L$ the nonlinear phase shift
is {\em deterministic\/} and $\phi_{\mathrm{NL}} = k_0 n_2 I_0 L$ for a wave
of intensity $I_0$, while in the case of diffuse multiple scattering
$\phi_{\mathrm{NL}}$ is {\em random\/} with the average value
$\left< \phi_{\mathrm{NL}} \right> \simeq k_0 n_2 I_0 (L^2/\ell)$ and variance
$\left< \phi_{\mathrm{NL}}^2 \right> -
\left< \phi_{\mathrm{NL}} \right>^2 \simeq p$.
The threshold of the speckle pattern instability $p \simeq 1$ is then simply
the point where the sample-to-sample fluctuations of the nonlinear phase
shift become of order unity.

To conclude this subsection, we would like to comment on the recent results on
modulation instability (MI) of incoherent light beams in homogeneous nonlinear media \cite{soljacic00}.
In this case, a 2D speckle pattern (speckled beam) with a controlled degree of spatial and temporal coherence is
`prepared' by sending a coherent laser beam on a rotating diffuser. The beam is then incident
on a homogeneous nonlinear medium (inorganic photorefractive crystal).
Pattern formation and `optical turbulence' is observed if the intensity of the beam exceeds a specific
threshold, determined by the degree of coherence of the beam (coherent MI has no threshold).
Although the study presented in this paper is also concerned with
the instability of speckle patterns in nonlinear media, it
differs essentially from that of Ref.\ \cite{soljacic00}
 at least in the following important aspects:
(a) the initially coherent wave loses its spatial coherence
\textit{inside} the medium, due to the multiple scattering on heterogeneities of the refractive index,
(b) the temporal coherence of the
incident light is perfect (in the case of incoherent MI \cite{soljacic00},
the coherence time of the incident beam is much shorter than the response time of the medium
$\tau_{\mathrm{NL}}$),
and (c) the incident light beam is completely destroyed by the scattering after a distance 
$\sim \ell$  and light propagation is diffusive in the bulk of the sample.
In addition, diffuse multiple scattering of light results in a distributed feedback mechanism,
absent in the case of MI.

\subsection{Path-Integral Picture}
\label{ssecpath}

A mathematically rigorous formulation of the heuristic analysis presented
in the previous subsection has been performed in Ref.\ \cite{skip00}.
In addition to the condition of weak nonlinearity (\ref{weak}) we anticipate that
the expected spontaneous fluctuations of the speckle pattern beyond the instability
threshold will be slow enough or, more precisely, that the typical time
of the speckle pattern dynamics will be much longer than the typical time
between two successive scattering events $\ell/c$.

For the sake of the mathematical simplicity, we consider a plane mono\-chro\-ma\-tic
wave incident on the stationary [$\delta \varepsilon(\vec{r})$ is time-independent]
semi-infinite disordered medium ($L \rightarrow \infty$)
with a finite (macroscopic) absorption length $L_{\mathrm{a}} \gg \ell$
and a local ($a_{\mathrm{NL}} = 0$) instantaneous ($\tau_{\mathrm{NL}} = 0$)
Kerr nonlinearity. It appears that
instead of replacing the intensity correlation function
$\left< \delta I(\vec{r}_1, t) \delta I(\vec{r}_2, t + \Delta t) \right>$ in
Eq.\ (\ref{vardi}) by its `linear' value (and thus constructing a sort of
perturbation theory), one can obtain a self-consistent
description of the speckle instability.
The analysis \cite{skip00} leads to a self-consistent equation for
$\beta = -\ln g_1$, where
$g_1(\tau) = \left< E(t) E^*(t+\tau) \right>/ \left< \left| E(t) \right|^2 \right>$
is the time autocorrelation function of diffusely reflected wave,
and $g_1$ denotes $\lim_{\tau \rightarrow \infty} g_1(\tau)$:
\begin{eqnarray}
\exp\left(-\beta \right) = F\left( \beta \right),
\label{self}
\end{eqnarray}
where $F(\beta)$ is a known monotonically decaying function and $F(0) = 1$.

\begin{figure}
\begin{center}
\includegraphics[width=0.45\textwidth]{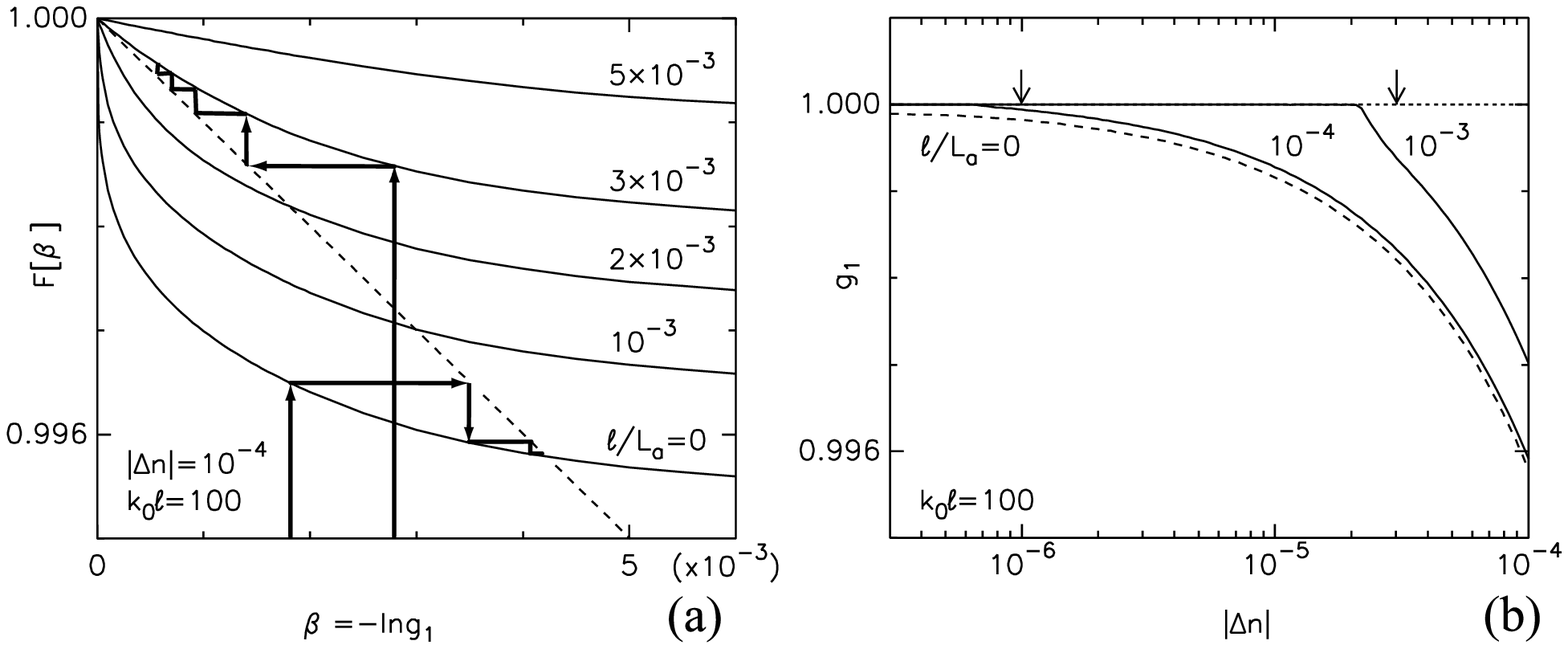}
\caption{\label{fig3}
(a) Graphical solution of Eq.\ (\ref{self}) at $k_0 \ell = 100$,
$\left| \Delta n \right| = 10^{-4}$ and the values of
$\ell/L_{\mathrm{a}}$ indicated near each curve. The solid lines
show $F(\beta)$, the dashed line is $\exp(-\beta)$.
Thick solid lines with arrows illustrate the iterative solution of
Eq.\ (\ref{self}).
(b) Bifurcation diagram of the speckle pattern in a semi-infinite
disordered medium for $k_0 \ell = 100$
and the values of $\ell/L_{\mathrm{a}}$ indicated near the each curve.
The threshold values of $\left| \Delta n \right|$, following from the
condition $p = 1$ at $\ell/L_{\mathrm{a}} > 0$, are indicated by arrows.
In the absence of
absorption, $\ell/L_{\mathrm{a}} = 0$ (dashed line), there is no threshold and
the speckle pattern is unstable at any (even infinitely small\,!)
$\left| \Delta n \right|$. The figures are from Ref.\ [16].}
\end{center}
\end{figure}

Fig.\ \ref{fig3}(a) illustrates the graphical solution of
Eq.\ (\ref{self}) for some realistic values of parameters.
At strong enough absorption [two upper curves in Fig.\ \ref{fig3}(a)], a unique
solution $\beta = \beta_1 = 0$ exists, corresponding to
$g_1 = \exp(-\beta) = 1$. This means that the time autocorrelation function
of diffusely reflected wave $g_1(\tau)$
does not decay with time, remaining equal to $1$
for all $\tau$ (even for $\tau \rightarrow \infty$).
The speckle pattern is therefore stationary
(i.e. does not change with time).
In contrast, for weak absorption [three lower curves in Fig.\ \ref{fig3}(a)]
a second solution $\beta = \beta_2 > 0$ arises, corresponding to
$g_1 = \exp(-\beta) < 1$. Eq.\ (\ref{self}) now has two fixed points:
$\beta = \beta_1 = 0$ and $\beta = \beta_2 > 0$. It is seen from Fig.\ \ref {fig3}(a)
(see thick solid lines with arrows)
that an iterative solution of Eq.\ (\ref{self}) starting at some $\beta > 0$
converges to $\beta = \beta_1 = 0$ for strong enough absorption
and to $\beta = \beta_2 > 0$ for weak absorption.
$\beta = \beta_2  > 0$
is therefore a stationary point of Eq.\ (\ref{self}) in the latter case,
corresponding to the physically realizable solution
beyond the instability threshold.
$\beta > 0$ and $g_1 < 1$ signifies
that the time autocorrelation function of diffusely
reflected wave $g_1(\tau)$ decays with $\tau$ despite our assumption
of stationary disorder [time-independent $\delta \varepsilon(\vec{r})$],
although the present analysis
does not allow us to estimate the characteristic time scale of this decrease.
Decaying $g_1(\tau)$ corresponds to an {\em unstable\/} (i.e. time-varying)
speckle pattern. Note that in contrast to the
fluctuations of the speckle pattern due to the motion of scattering centers,
employed in DWS to study the dynamics of the medium \cite{maret87},
the speckle dynamics in a stationary nonlinear random
medium is {\em spontaneous,} i.e. it does not originate from the motion of
scattering centers (since the latter are immobile) but is intrinsic for the
underlying nonlinear wave equation (\ref{weq}).
Spontaneous speckle dynamics is irreversible and hence the instability threshold
$p \simeq 1$ is the point where the time-reversal symmetry is spontaneously
broken for the multiple-scattered waves.

Fig.\ \ref{fig3}(b) shows the time autocorrelation function
$g_1 = \exp(-\beta)$ obtained by solving Eq.\ (\ref{self}) at
$k_0 \ell = 100$ and several values of $\ell/L_{\mathrm{a}}$.
The transition from the stable ($g_1 = 1$) to unstable ($g_1 < 1$) regime is
clearly seen when $\ell/L_{\mathrm{a}} > 0$. At $\ell/L_{\mathrm{a}} = 0$
(no absorption), we find that $g_1 < 1$ at any, even infinitely small
$\left| \Delta n \right|$.
The border between stable ($g_1 = 1$) and unstable ($g_1 < 1$) regimes can be
found analytically by requiring $(\partial/\partial \beta) F(\beta) = -1$
at $\beta = 0$.
This yields $p \simeq 1$ as the instability threshold, where
the bifurcation parameter $p$ is defined by Eq.\ (\ref{p}) with
$L$ replaced by $L_{\mathrm{a}}$.
The threshold values of $\left| \Delta n \right|$ following from the condition
$p = 1$ are shown in Fig.\ \ref{fig3}(b) by arrows.

\subsection{Langevin Description}
\label{sseclang}

Although the path-integral approach, sketched in the two previous subsections,
allows us to predict the instability of the multiple-scattering
speckle pattern in a nonlinear disordered medium and even to calculate 
$g_1(\tau)$ in the limit $\tau \rightarrow \infty$, it does not permit to
estimate the time scale of spontaneous intensity fluctuations beyond the instability
threshold.
Meanwhile, this issue is of primary importance in view of the
possible experimental observation of the instability phenomenon.
In this subsection, we show that
the spontaneous speckle dynamics can be studied by
using the Langevin approach \cite{skip02a,skip02b}.

As in the previous subsections, we assume the nonlinearity to be local
($a_{\mathrm{NL}} = 0$), but consider arbitrary nonlinearity response time
$\tau_{\mathrm{NL}}$. The nonlinear part of the dielectric constant is assumed to
be governed by Eq.\ (\ref{debye}).
In addition, we neglect the absorption (assuming $L_{\mathrm{a}} \gg L$)
and restrict our consideration to the case when
the development of the speckle pattern instability is dominated by the
long-range intensity correlations.
This requires the sample size to be large enough [$L \gg k_0 \ell^2$,
see Eq.\ (\ref{p}) and its accompanying discussion]
and the speckle dynamics to be slow
(time scale of spontaneous intensity fluctuations
$\tau \gg T_{\mathrm{D}} [(k_0 \ell^2)/L]^2$).
The latter condition follows from Eq.\ (\ref{varnl2}), where the
limits of integration over $\Delta s$ should be set to $\mp c \tau$, if
$c \tau \ll s_0$ (or, equivalently, $\tau \ll T_{\mathrm{D}}$), since
the correlation has simply no time to establish beyond these limits due to the
finite speed of wave propagation.
Integration in Eq.\ (\ref{varnl2}) then yields
\begin{eqnarray}
\left< \Delta \phi_{\mathrm{NL}}^2 \right> \simeq
\Delta n^2 \left( \frac{L}{\ell} \right)^2
\left[ k_0 \ell + \frac{L}{\ell} \left( \frac{\tau}{T_{\mathrm{D}}} \right)^{1/2}
\right]
\left< \Delta \phi_{\mathrm{L}}^2 \right>,\nonumber \\
\label{varnl4}
\end{eqnarray}
where the first term in square brackets [the same as in Eq.\ (\ref{p})] results
from the short-range correlation (\ref{short}), while the second term is due to
the long-range one (\ref{long}). Requiring that the second term dominates the
first, we obtain $\tau \gg T_{\mathrm{D}} [(k_0 \ell^2)/L]^2$ as stated above.

\begin{figure}
\begin{center}
\includegraphics[width=0.45\textwidth]{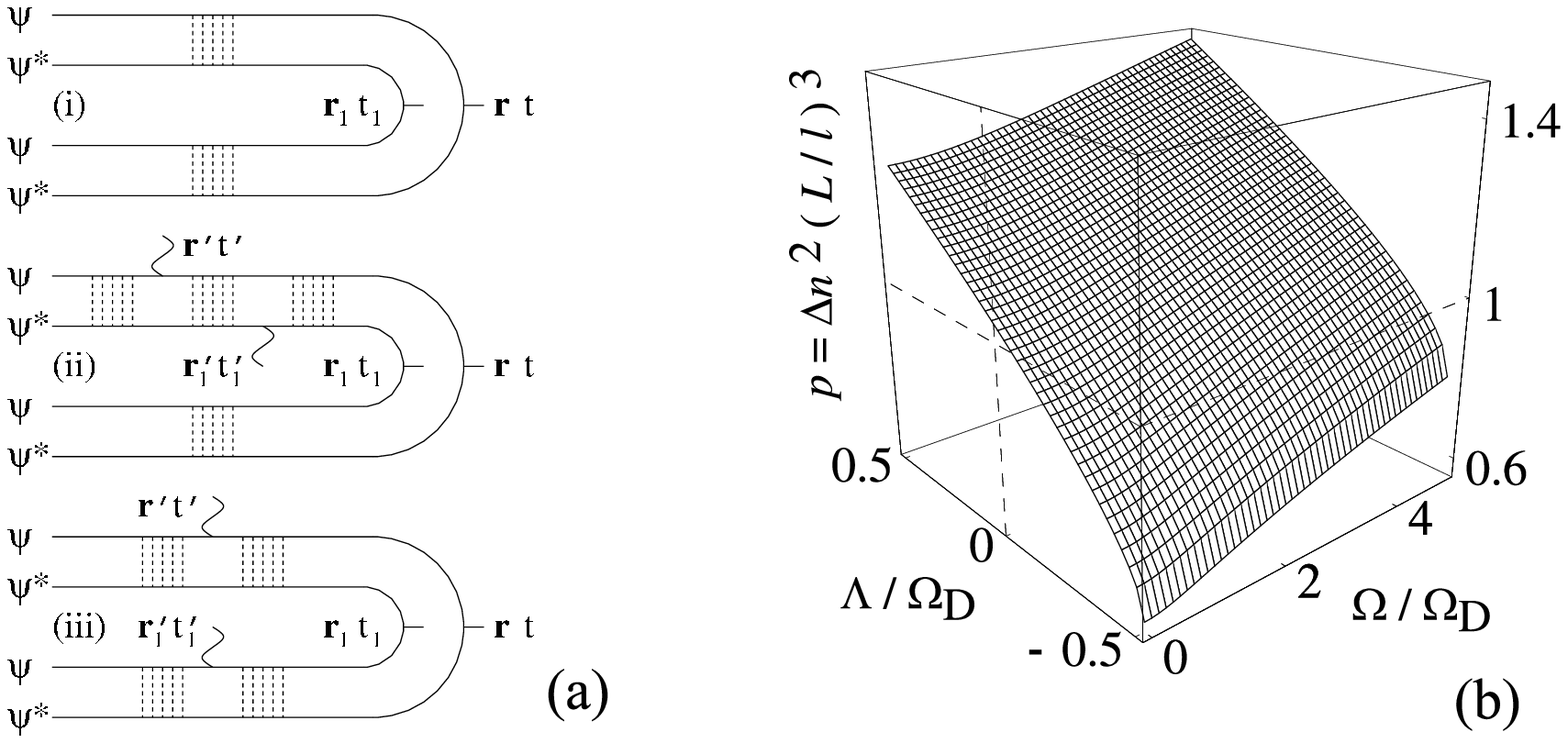}
\caption{\label{fig4}
(a) Diagrams contributing to the correlation functions of Langevin currents
$\vec{j}_{\mathrm{ext}} (\vec{r}, t)$ [diagram (i)] and random response functions
$\vec{q} (\vec{r}, \vec{r}^{\,\prime}, \Delta t)$
[diagrams (ii) and (iii)]. The diagrams (ii) and (iii) are obtained by
the functional differentiation of the diagram (i) with respect to
the dielectric constant of the disordered medium. Wavy lines in the diagrams
(ii) and (iii) denote $k_0^2$ factors.
(b) Surface describing the stability of the multiple-scattering speckle pattern
in a disordered medium with fast nonlinearity
($\Omega_{\mathrm{D}} \tau_{\mathrm{NL}} \ll 1$). For given frequency
$\Omega$ and bifurcation parameter $p$, the surface shown in the
figure allows one to determine the value of the Lyapunov exponent $\Lambda$.
If $\Lambda > 0$, the speckle pattern is unstable with respect to excitations
at frequency $\Omega$. The figures are from Ref.\ [18].}
\end{center}
\end{figure}

The Langevin equation for the intensity fluctuation $\delta I(\vec{r}, t)$
reads \cite{zyuzin87}:
\begin{eqnarray}
\frac{\partial}{\partial t} \delta I(\vec{r}, t)  - D \nabla^2  \delta I(\vec{r}, t) =
- \vec{\nabla} \cdot {\vec j}_{\mathrm{ext}}(\vec{r}, t),
\label{langevin}
\end{eqnarray}
where ${\vec j}_{\mathrm{ext}}(\vec{r}, t)$ are random external Langevin currents that
have zero mean and a correlation function
$\left< j_{\mathrm{ext}}^{(i)} (\vec{r}, t) j_{\mathrm{ext}}^{(j)} (\vec{r}_1, t_1)
\right>$ given by the diagram (i) of Fig.\ \ref{fig4}(a).
Being a `fingerprint' of disorder, the Langevin currents will be modified if we
modify the dielectric constant of the medium. In our case, the linear part
of the dielectric constant is fixed [$\delta \varepsilon(\vec{r})$
in Eq.\ (\ref{weq}) is time-independent], but its nonlinear part
$\Delta \varepsilon_{\mathrm{NL}}(\vec{r}, t)$ can vary with time, and
it is relatively easy to obtain the following dynamic equation
for $\vec{j}_{\mathrm{ext}} (\vec{r}, t)$ \cite{skip02a,skip02b}:
\begin{eqnarray}
&&\frac{\partial}{\partial t} \vec{j}_{\mathrm{ext}} (\vec{r}, t) =
\int_V d^3 \vec{r}^{\, \prime} \int_{0}^{\infty} d \Delta t \,
\nonumber \\
&&\vec{q}( \vec{r}, \vec{r}^{\, \prime}, \Delta t) \;
\frac{\partial}{\partial t}
\Delta \varepsilon_{\mathrm{NL}}(\vec{r}^{\, \prime}, t - \Delta t),
\label{djnl}
\end{eqnarray}
where $\vec{q}( \vec{r}, \vec{r}^{\, \prime}, \Delta t)$ is a random response function
with zero mean and the correlation function
$\left< q^{(i)} (\vec{r}, \vec{r}^{\,\prime}, \Delta t)
q^{(j)*} (\vec{r}_1, \vec{r}_1^{\,\prime}, \Delta t_1) \right>$
given by a sum of the
diagrams (ii) and (iii) of Fig.\ \ref{fig4}(a).
Eqs.\ (\ref{langevin}) and (\ref{djnl}) together with Eq.\ (\ref{debye}) for
$\Delta \varepsilon_{\mathrm{NL}}$ form a self-consistent system of equations
for the stability analysis of the speckle pattern.

Consider now an infinitesimal periodic excitation of the stationary speckle pattern:
$\delta I (\vec{r}, t) =  \delta I (\vec{r}, \alpha)$ $\exp( \alpha t )$,
where $\alpha = i \Omega + \Lambda \ne 0$ and $\Omega > 0$.
Such an excitation can be either damped or
amplified, depending on the sign of the Lyapunov exponent $\Lambda$.
The value of $\Lambda$ is determined by two competing processes: on the one hand,
diffusion tends to smear the excitation out, while on the other hand, the
distributed feedback sustains its existence.
The mathematical description of this competition is provided by
Eqs.\ (\ref{langevin}), (\ref{djnl}), and (\ref{debye})
that after the substitution of
$\delta I (\vec{r}, t) =  \delta I (\vec{r}, \alpha) \exp( i \alpha t )$
[and similarly for $\vec{j}_{\mathrm{ext}} (\vec{r}, t)$ and
$\Delta \varepsilon_{\mathrm{NL}}(\vec{r}, t)$] lead
to the following equation for $p$, $\Omega$, and $\Lambda$ \cite{skip02a,skip02b}:
\begin{eqnarray}
p \simeq F_1 \left( \Omega/\Omega_{\mathrm{D}}, \Lambda/\Omega_{\mathrm{D}} \right)
F_2 \left( \Omega \tau_{\mathrm{NL}}, \Lambda \tau_{\mathrm{NL}} \right),
\label{final}
\end{eqnarray}
where $\Omega_{\mathrm{D}} = 1/T_{\mathrm{D}} = D/L^2$ and
the function $F_1$ is shown in Fig.\ \ref{fig4}(b) for the case of the sample
with open boundaries, while
$F_2(x, y) = 1 + x^2 + y^2 + 2 y$.

Let us first consider the fast nonlinearity, assuming
$\tau_{\mathrm{NL}} \Omega_{\mathrm{D}} \ll 1$ \cite{skip02a}. In this case, the nonlinear response
takes much less time than the typical perturbation $\delta I(\vec{r}, \alpha)$ needs
to extend throughout the disordered sample and we can set
$F_2 \left( \Omega \tau_{\mathrm{NL}}, \Lambda \tau_{\mathrm{NL}} \right)
\simeq 1$ in Eq.\ (\ref{final}). The stability of the speckle pattern with respect
to periodic excitations is then described by Fig.\ \ref{fig4}(b).
For a given frequency $\Omega$, the sign of the Lyapunov exponent $\Lambda$ depends
on the value of $p$.
Excitations at frequencies $\Omega$ corresponding to $\Lambda < 0$ are damped exponentially and thus
soon disappear. In contrast, excitations at frequencies $\Omega$ corresponding to $\Lambda > 0$
are exponentially amplified, which signifies the instability of
the speckle pattern with respect to excitations at such frequencies.
Noting that $\Lambda$  is always negative for $p < 1$, we conclude that
all excitation are damped in this case and the
speckle pattern is absolutely stable. In an experiment, any spontaneous excitation of the static speckle
pattern will be suppressed and the speckle pattern will be independent of time:
$\delta I(\vec{r}, t) = \delta I(\vec{r})$, as in the linear case.
When $p > 1$, an interval of frequencies $0 < \Omega <
\Omega_{\mathrm{max}}$ starts to open up with $\Lambda > 0$.
The speckle pattern thus becomes unstable with respect to
excitations at low frequencies. In an experiment,
any spontaneous excitation of the static
speckle pattern at frequency $\Omega \in (0, \Omega_{\mathrm{max}})$
will be amplified and one will observe a time-varying speckle pattern
$\delta I(\vec{r}, t)$.
Note that the absolute instability threshold $p \simeq 1$ agrees with the results of
Secs.\ \ref{ssecheur} and \ref{ssecpath}, while completely different calculation
techniques have been applied in the three cases.

\begin{figure}
\begin{center}
\includegraphics[width=0.45\textwidth]{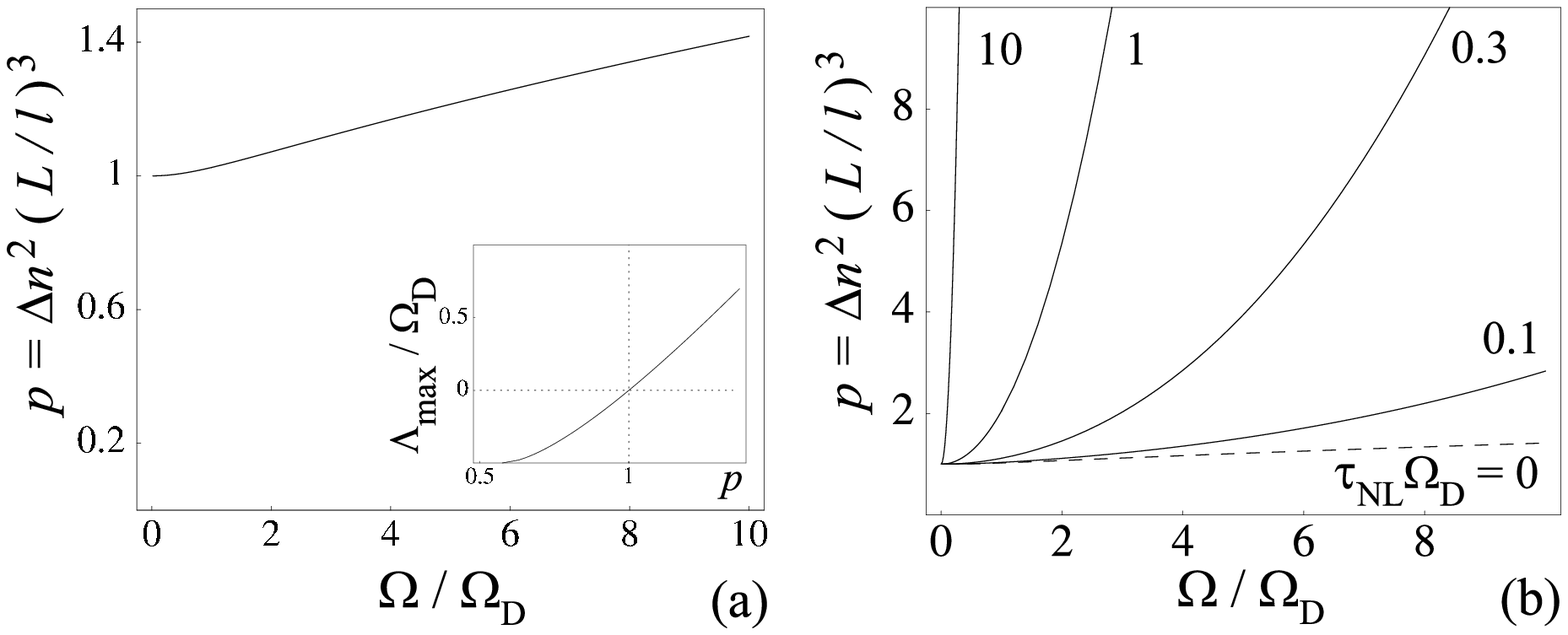}
\caption{\label{fig5}
(a) Frequency-dependent instability threshold for a disordered sample with
instantaneous nonlinear response and
open boundaries. The inset shows the
maximal Lyapunov exponent as a function of the bifurcation parameter
$p$. (b) The same as (a) but for a sample with a
nonzero nonlinearity response time $\tau_{\mathrm{NL}}$.
The dashed line shows the result corresponding
to $\tau_{\mathrm{NL}} \Omega_{\mathrm{D}} = 0$ [the same as the solid line
in the panel (a)]. The figures are from Ref.\ [18] (a) and [19] (b).}
\end{center}
\end{figure}

Fig.\ \ref{fig5}(a) shows the frequency-dependent instability threshold for
the disordered sample with open boundaries and fast nonlinear response.
Detailed analysis \cite{skip02a} of Eq.\ (\ref{final}) shows that
the frequency-dependent
threshold value of $p$ scales as $1 + C_1 (\Omega/\Omega_{\mathrm{D}})^2$
(where $C_1 \sim 1$ is a numerical constant) at
$\Omega \ll \Omega_{\mathrm{D}}$
and as $(\Omega/\Omega_{\mathrm{D}})^{1/2}$
at $\Omega \gg \Omega_{\mathrm{D}}$. The latter result can also be obtained from
Eq.\ (\ref{varnl4}) by setting $\tau \sim 1/\Omega$ and applying
$\left< \Delta \phi_{\mathrm{NL}}^2 \right> > \left< \Delta \phi_{\mathrm{L}}^2 \right>$
as the instability condition.
For a given value of $p > 1$, the maximum excited frequency is
$\Omega_{\mathrm{max}} \sim \Omega_{\mathrm{D}} (p - 1)^{1/2}$ for
$p - 1 \ll 1$ and $\Omega_{\mathrm{max}} \sim \Omega_{\mathrm{D}} p^{2}$ for
$p -1 > 1$.

Fig.\ \ref{fig5}(b) illustrates the effect of noninstantaneous nonlinearity on the
frequency-dependent instability threshold for several values of
$\tau_{\mathrm{NL}} \Omega_{\mathrm{D}}$ \cite{skip02b}.
Obviously, slow nonlinear response of
the medium rises the instability threshold for high-frequency excitations, while
the absolute instability threshold $p \simeq 1$ remains the same, independent
of the nonlinearity response time $\tau_{\mathrm{NL}}$. The reason for this is that
just above $p = 1$ the speckle pattern becomes unstable with respect to
excitations at very low frequencies, for which $\Omega \ll 1/\tau_{\mathrm{NL}}$
is always fulfilled and which, therefore, are not sensitive to the value of
$\tau_{\mathrm{NL}}$ [mathematically,
$F_2 ( \Omega \tau_{\mathrm{NL}}, \Lambda \tau_{\mathrm{NL}} ) \simeq 1$
in Eq.\ (\ref{final})].
In the limit of slow nonlinearity
($\tau_{\mathrm{NL}} \Omega_{\mathrm{D}} \gg 1$) and for $1 < p - 1 \ll 1$ we can set
$F_1 \left( \Omega/\Omega_{\mathrm{D}}, \Lambda/\Omega_{\mathrm{D}} \right) \simeq 1$
in Eq.\ (\ref{final}) and find the threshold value of $p$ to scale as
$1 + (\Omega \tau_{\mathrm{NL}})^2$ and
$\Omega_{\mathrm{max}} \sim \tau_{\mathrm{NL}}^{-1} (p - 1)^{1/2}$.

As follows from the above analysis, a continuous low-frequency spectrum
of frequencies $(0, \Omega_{\mathrm{max}})$
is excited at $p > 1$. In addition, the Lyapunov exponent $\Lambda$ appears to decrease
monotonically with $\Omega$, favoring no specific frequency $\Omega$ just above the threshold
\cite{skip02b}.
This allows us to hypothesize that at $p = 1$ the speckle pattern undergoes
a transition from a stationary to chaotic state.
Such a behavior should be contrasted from the `route to chaos' through a sequence
of bifurcations, characteristic of many nonlinear and, in particular,
optical  systems \cite{ikeda80}--\cite{voron99}.

\section{Conclusion}
\label{secconcl}

The present paper reviews some of the recent theoretical developments in the field of
multiple wave scattering in nonlinear disordered media. To be specific, we consider optical waves and
restrict ourselves to the case of Kerr nonlinearity. Assuming that the nonlinearity is weak,
we derive the expressions for the angular correlation functions and the coherent backscattering cone
in a nonlinear disordered medium (Sec.\ \ref{secangular}). In both transmission and reflection,
the short-range angular correlation functions of intensity fluctuations for two waves with
different amplitudes ($A$ and $A^{\prime} \rightarrow 0$) appear to be given by the
same expressions [Eqs.\ (\ref{ca2}) and (\ref{ca3}), respectively]
as the angular correlation functions for waves at two different frequencies
($\omega$ and $\omega^{\prime} = \omega - \Delta \omega$) in a linear medium, with $\Delta \omega$ replaced
by $2 \ell c/(3 \xi^2)$, where $\xi$ is a new
{\it nonlinear} characteristic length defined by Eq.\ (\ref{xi}).
The coherent backscattering cone is not affected by the nonlinearity, as long as the
nonlinear coefficient $\varepsilon_2$ in Eq.\ (\ref{weq}) is purely real. If $\varepsilon_2$ has an
imaginary part (which corresponds to the nonlinear absorption), the line shape of the cone is given
by the same expression (\ref{cbs}) as in an absorbing linear medium,
where the linear macroscopic absorption length
$L_{\mathrm{a}}$ should be replaced by the {\it generalized} absorption length $L_{\mathrm{a}}^{\mathrm{NL}}$
defined by Eq.\ (\ref{lastar}).

For the nonlinearity strength exceeding a threshold $p \simeq 1$ [with the bifurcation parameter
$p$ given by Eq.\ (\ref{p})],
we predict a new phenomenon ---
temporal instability of the multiple-scattering speckle pattern --- to take place
(Sec.\ \ref{secinst}).
The instability is due to a combined
effect of the nonlinear self-phase modulation and the distributed feedback mechanism provided by multiple scattering
and should manifest itself in spontaneous fluctuations of the speckle pattern with time.
Since the spontaneous dynamics of the speckle pattern is irreversible, the time-reversal symmetry is
spontaneously broken when $p$ surpasses 1.
The important feature of our result is the extensive nature of the instability threshold,
leading to an interesting possibility of obtaining unstable regimes even at
very weak nonlinearities, provided that the disordered sample is large
enough. To study the dynamics of multiple-scattering speckle patterns beyond the
instability threshold, we generalize the Langevin description of wave diffusion in disordered media
(Sec.\ \ref{sseclang}).
Explicit expressions for the characteristic time scale of spontaneous intensity fluctuations are derived
with account for the noninstantaneous nature of the nonlinearity. The results of this study allow us
to hypothesize that the dynamics of the speckle pattern may become chaotic immediately beyond the
instability threshold, and that the cascade of bifurcations, typical for chaotic transitions in many
known nonlinear systems \cite{ikeda80}--\cite{voron99}, might not be present in the considered case
of diffuse waves.

Finally, we discuss the experimental implications of our results.
With common nonlinear materials, such as, e.g., carbon disulfide, $n_2 \simeq 3 \times 10^{-14}$
cm$^2$/W can be realized \cite{boyd02}. At $I \simeq 1$ MW/cm$^2$ this yields
$\Delta n \simeq 3 \times 10^{-8}$ and the characteristic length $\xi$ defined in Eq.\ (\ref{xi})
is $\xi \simeq 2$ cm for $\ell \simeq 100$ $\mu$m and $\lambda \simeq 0.5$ $\mu$m.
As follows from Eq.\ (\ref{ca5}), $\xi \lesssim L$ is required to observe a sizeable effect of the
nonlinearity on the angular correlation function of transmitted light, and hence using a
2 cm-thick disordered sample should suffice to make the effect of nonlinearity measurable in, e.g., a dense
suspension of carbon disulfide particles.
In contrast, $\xi \lesssim \ell$ is necessary to observe the effect of the nonlinearity on the
angular correlation function in the reflection geometry. This requires a stronger nonlinearity.
In nematic liquid crystals, for example, $\Delta n$ up to $ 0.1$ is achievable (see, e.g., Ref.\ \cite{muenster97}).
Taking $\Delta n = 10^{-3}$ and $\ell \sim 1$ mm \cite{kao96}, we get
$\xi \sim 1$ mm $\sim \ell$.
In the case of the coherent backscattering cone, very accurate experimental techniques developed
recently to measure the angular dependence of backscattered light \cite{wiersma95}
give a hope that the effect of the nonlinearity can be observable without any particular problem.
Finally, observation of the temporal instability of multiple-scattering speckle patterns, predicted
in Sec.\ \ref{secinst}, will require, first of all, large enough sample size $L$ and as low absorption
as possible. In the absence of absorption (or for the macroscopic absorption length
$L_{\mathrm{a}} \gtrsim L$),  $\Delta n_{\mathrm{NL}} \sim 10^{-2}$ (realistic in nematic liquid
crystals \cite{muenster97}) and $L/\ell \sim 20$ will suffice to
get $p \simeq 1$ and reach the instability threshold.
We note that in liquid crystals, the nonlinearity is due to the reorientation of molecules under the
influence of the electric field of the electromagnetic wave and hence is essentially
noninstantaneous. This emphasizes the importance of including the noninstantaneous
nature of nonlinear response in our analysis (Sec.\ \ref{sseclang}).

\end{document}